\documentclass[12pt,a4]{article}

\usepackage[dvips]{graphicx}
\begin{document}
\setlength{\baselineskip}{2em}
%\bibliographystyle{prsty}
%\draft
%\preprint{Proceedings of Analytical Study of Quantum Information and
%          Related Fields, Jan. 25 to 27, 2000 in Kyoto}

\title{Correlations of measurement information and noise in 
quantum measurements with finite resolution}

\author{Holger F. Hofmann and Takayoshi Kobayashi \\
Department of Physics, Faculty of Science, University of Tokyo,\\
7-3-1 Hongo, Bunkyo-ku, Tokyo113-0033, Japan\\[0.1cm]
Akira Furusawa\\ Nikon Corporation, R\&D Headquarters,\\
Nishi-Ohi, Shinagawa-ku, Tokyo 140-8601, Japan}

\date{\today}

\maketitle

\begin{abstract}
The original purpose of measurements is to provide us with information 
about a previously unknown physical property of the system observed.
In the Hilbert space formalism of quantum mechanics, this physical 
meaning of measurement information is not immediately apparent. 
In order to study the relationship between the Hilbert space coherence
of the quantum state and the measurement information obtained in the 
laboratory, we introduce a generalized measurement postulate for 
finite resolution measurements. With this measurement model, correlations
between non-commuting observables can be investigated.
These experimentally accessible correlations reveal nonclassical features
in their dependence on the operator ordering, reflecting the particular 
measurement context by which they are determined.
\end{abstract}

\section{Introduction}
The interpretation of quantum mechanics has been controversial from
the very beginning, and even one hundred years after Planck's quantum 
hypothesis, the physical reality behind the formalism is still being
debated \cite{PT99}. At the heart of this confusion is the issue of
quantum measurements \cite{Whe83}. The original definitions of physical
quantities were based on the expectation that the quantities observed
in measurements are effectively identical to the fundamental mathematical 
elements of the theory. The implicit assumption behind this expectation is
that knowledge and fact are independent and seperable. An ideal measurement
simply provides information about facts without changing the facts.
Reality would then be defined by an infinitely precise set of observable
quantities. However, this assumption had to be abandoned in order to 
describe atomic and optical phenomena. Starting with Planck's quantum
hypothesis, quantum theory completely abandoned the constraints of 
classical physics. In order to justify this radical 
departure from these highly successful concepts, Bohr and Heisenberg argued 
that no reality needs to be attributed to properties which are not being 
observed. In the famous Bohr-Einstein dialogue, the decisive argument was 
provided by the uncertainty principle which places a fundamental restriction 
on our ability to know observable properties. 

\begin{figure}
\hspace{1.25 cm}
\begin{picture}(345,200)
%\put(0,0){\framebox(345,200){}}
\put(0,0){\framebox(160,200){}}
\put(0,150){\makebox(160,30){\Large Physical principles}}
\put(0,120){\makebox(160,30){\Large Observable properties}}
\put(0,100){\makebox(160,30){\Large in space and time}}
\put(0,60){\makebox(160,30){\Large Uncertainty relations}}
\put(0,40){\makebox(160,30){\Large $\Downarrow$}}
\put(0,20){\makebox(160,30){\Large Quantum Noise}}
\put(185,0){\framebox(160,200){}}
\put(185,160){\makebox(160,25){\Large Mathematical}}
\put(185,140){\makebox(160,25){\Large principles}}
\put(185,120){\makebox(160,30){\Large Abstract formalism}}
\put(185,100){\makebox(160,30){\Large in Hilbert space}}
\put(185,60){\makebox(160,30){\Large States and Operators}}
\put(185,40){\makebox(160,30){\Large $\Downarrow$}}
\put(185,20){\makebox(160,30){\Large Quantum Information}}
\end{picture}
\caption{\label{dual} Illustration of the dualism between the mathematical 
principles and the physical principles of quantum mechanics.}
\end{figure}
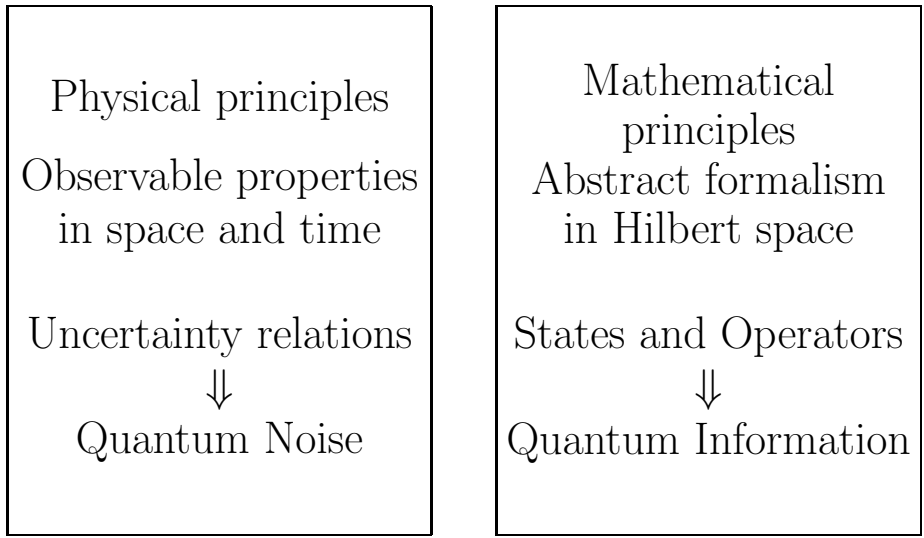

At the same time, a highly precise mathematical description of quantum 
mechanics emerged. This mathematical description introduces the concept
of probability amplitudes, which combine the epistemological
nature of probabilities with the deterministic concept of interference. 
It is this Hilbert space formalism which has firmly established quantum 
mechanics in 20th century physics. Nevertheless interpretational problems 
remain because the physical meaning of the Hilbert space formalism is 
purely statistical. The quantum state of a single system cannot be measured.
This situation gives rise to a dualism between the mathematical
description and the physical meaning in quantum mechanics \cite{Hei58} 
which is illustrated in figure \ref{dual}. 
For a long time, this dualism was not recognized as experiments mostly
revealed statistical averages of only a few observable properties.
As modern technology allows a more detailed investigation of quantum 
phenomena, however, the interpretational problems associated with a 
theory that is not based upon directly observable facts have reemerged.
It may therefore be necessary to investigate the role of measurement
in quantum mechanics with respect to the possible manipulations of single
quantum objects in order to clarify the physical meaning of quantum states.
In the light of the present discussion on the usefulness of quantum effects 
for computation and data transmission, a clarification of the process by
which information is extracted from a quantum system may also
provide a better understanding of the technical requirements for
quantum information processing.

In the following, the relationship between the measurement information obtained
at finite resolution and the noise introduced by the measurement interaction
is investigated. It is shown that the noise is strongly correlated with the
measurement result, even if the eigenvalues of the observed property would
not permit such correlations. These correlations represent operator ordering
dependent properties which emerge only in the regime where the measurement
resolution is high enough to reveal quantization, yet low enough to permit
coherence. Thus, finite resolution measurements reveal more about the physical
properties of quantum systems than precise measurements.
Hopefully, the investigation of these nonclassical correlations will help 
to bridge the gap between the mathematical principles and the physical 
principles of quantum mechanics by providing a physical 
interpretation of quantum coherence and operator ordering.

\section{Uncertainty in quantum measurements}
The connection between the mathematical formalism and the physical
observables is established by the measurement. The squared
probability amplitudes of Hilbert space formalism can then be interpreted  
as real probabilities. The probabilities can be summarized in the 
density matrix $\hat{\rho}$ which may be expressed in terms of the 
eigenstates of $\hat{n}$ as
\begin{equation}
\hat{\rho}=\sum_{n,m} \rho_{nm} \mid n\rangle\langle m \mid.
\end{equation}
While the diagonal elements $\rho_{nn}$ clearly define the probability of
obtaining a measurement result of $n$ in a precise measurement of 
$\hat{n}$, the relationship between $\hat{n}$ and the off-diagonal
elements $\rho_{nm}$ with $n\neq m$ is less clear. The physical meaning 
of the off-diagonal elements only emerges in a transformation to a different
set of eigenstates, where $\rho_{nm}$ modifies the new probabilities as
an interference term. However, a precise measurement of $\hat{n}$ 
destroys this interference information, increasing the uncertainty in
the probability distributions associated with eigenstates of observables
other than $\hat{n}$ in accordance with the uncertainty principle.

Uncertainty requires the dissapearance of $\rho_{nm}$ before $n$ can
be distinguished from $m$. The information obtained about $\hat{n}$
therefore requires that the measurement interaction introduces 
decoherence in the off-diagonal elemnts $\rho_{nm}$. If the noise
does not reduce $\rho_{nm}$ to zero, then the resolution $\delta\!n$ is
not sufficient to completely distinguish $n$ from $m$. 
This situation is typical for optical quantum nondemolition
measurements of photon number \cite{Cav80,Lev86,Fri92}.
In such experiments, the information obtained about the photon number 
$\hat{n}$ is given by a measurement result $n_m$ and a resolution 
$\delta n$ which corresponds to the Gaussian uncertainty of the measurement. 
However, this reduced measurement resolution is not just a limitation but 
helps to preserve the coherence of the quantum state \cite{Imo85,Kit87}. 

Instead of selecting a well-defined photon number component 
$\mid n \rangle$, the measurement adjusts the statistical weight of each 
component $\mid n \rangle$ by a factor dependent
on the difference between $n_m$ and $n$. This effect of the measurement can be
represented by a generalized measurement operator 
$\hat{P}_{\delta\!n}(n_m)$ \cite{Hof20} given by
\begin{equation}
\hat{P}_{\delta n}(n_m)= (2\pi \delta\! n^2)^{-1/4}\; 
              \exp \left(-\frac{(n_m-\hat{n})^2}{4 \delta\! n^2}\right)
.
\end{equation}
For a given initial state $\hat{\rho_i}$,
the probability distribution $P(n_m)$ over measurement results $n_m$
and the density matrix $\hat{\rho}_f(n_m)$ after the measurement read
\begin{eqnarray}
\label{eq:prob}
P(n_m) &=& 
Tr\left\{\hat{P}_{\delta\!n}(n_m)\;\hat{\rho}(\mbox{in})\;
\hat{P}_{\delta\!n}(n_m) \right\}
\\
\hat{\rho}_f(n_m) &=& \frac{1}{P(n_m)}
\hat{P}_{\delta\!n}(n_m)\;\hat{\rho}(\mbox{in})\;\hat{P}_{\delta\!n}(n_m).
\end{eqnarray}
By describing the effects of a finite measurement resolution
$\delta n$ on all physical properties of the system
the generalized measurement postulate defined by the operator 
$\hat{P}_{\delta\!n}(n_m)$ provides an expression of the role
of uncertainty in quantum measurements.
It is then possible to quantify the changes in coherence associated with a
photon number measurement result $n_m$ in detail.

\section{Photon number measurement statistics}
The interference properties of the light field are described by the 
coherent amplitude operator $\hat{a}$. This operator is also referred to
as annihilation operator because its effect on photon number states is
given by
\begin{equation}
\hat{a}\mid n \rangle = \sqrt{n}\mid n-1 \rangle.
\end{equation}
However, the physical meaning of this photon number property is 
simply that the interference properties of the light field given
by the expectation value $\langle \hat{a} \rangle$ depend
on the density matrix elements $\rho_{nm}$ with $m=n-1$. 
Photon number information therefore destroys the interference 
properties by reducing $\langle \hat{a} \rangle$. 

\begin{figure}
\vspace{-1cm}
\hspace{1.25 cm}
\begin{picture}(345,250)
%\put(0,0){\framebox(345,250)}
\put(0,0){\makebox(345,250){
\includegraphics[width=12cm]{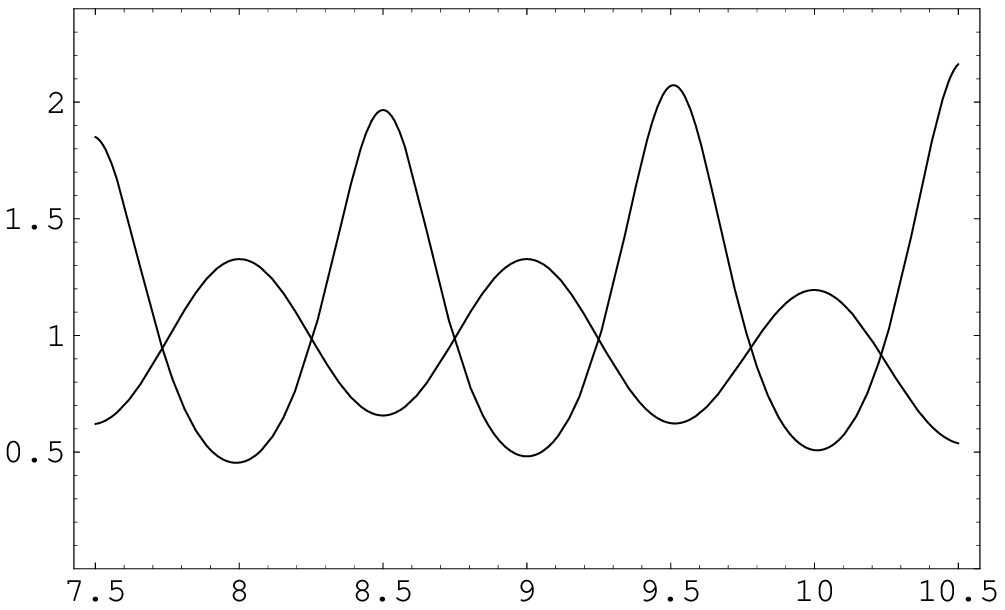}}}
\put(180,0){\makebox(20,20){\Large $n_m$}}
\put(78,42){\makebox(60,20)
     {\Large $\frac{P(n_m)}{P(n_m=9.25)}$}}
\put(114,69){\vector(1,1){15}}
\put(45,202){\makebox(70,20)
        {\Large $\frac{\langle \hat{a}\rangle_f(n_m)}
               {\langle \hat{a}\rangle_{f}(n_m=9.25)}$}}
\put(100,197){\vector(1,-1){20}}
\end{picture}
\caption{\label{detail}
Details of the quantum mechanical
modulations of measurement probability and coherence after the measurement
near $n_m=9$, normalized by the results at $n_m=9.25$.}
\end{figure}

For a coherent light field such as that emitted by a typical 
laser, the initial density matrix is given by 
\begin{equation}
\hat{\rho}_i=\sum_{n,m} \exp(-|\alpha|^2)
\frac{1}{\sqrt{n!m!}}(\alpha^*)^m(\alpha)^n \mid\! n\rangle\langle m\!\mid.
\end{equation}
This coherent state has an average amplitude of 
$\langle \hat{a} \rangle = \alpha$
and an average photon number of $\langle \hat{n} \rangle
= |\alpha|^2$.
The optical coherence of this state is maximal and any reduction 
in the photon number fluctuations leads to a reduction of the coherence
given by $\langle \hat{a} \rangle = \alpha$. 
By applying the general measurement postulate according to equation
(\ref{eq:prob}), one obtains the probability distribution for photon number
measurement results $n_m$ at a finite resolution of $\delta\! n$ as
\begin{equation}
P(n_m) = \frac{\exp(-|\alpha|^2)}{\sqrt{2\pi\delta\! n^2}}
      \sum_n \frac{|\alpha|^{2n}}{n!} 
             \exp\left(-\frac{(n-n_m)^2}{2\delta\! n^2}\right).
\end{equation}
This probability distribution is simply a sum of Gaussians centered around
the respective integer photon numbers. At low resolution (high $\delta\!n$),
the Gaussians merge to form a continuous probability distribution. At
high resolution (low $\delta\!n$), the Gaussians are completely separate.
However, coherence is only possible in the regions where neighbouring 
Gaussians overlap. This becomes clear if the coherence 
$\langle \hat{a} \rangle_f(n_m)$ after a measurement of $n_m$ is determined.
\begin{figure}
\begin{picture}(345,420)
%\put(0,0){\framebox(345,420)}
\put(90,380){\makebox(20,20){\Large (a)}}
\put(40,230){\makebox(300,180){
\includegraphics[width=10.2cm]{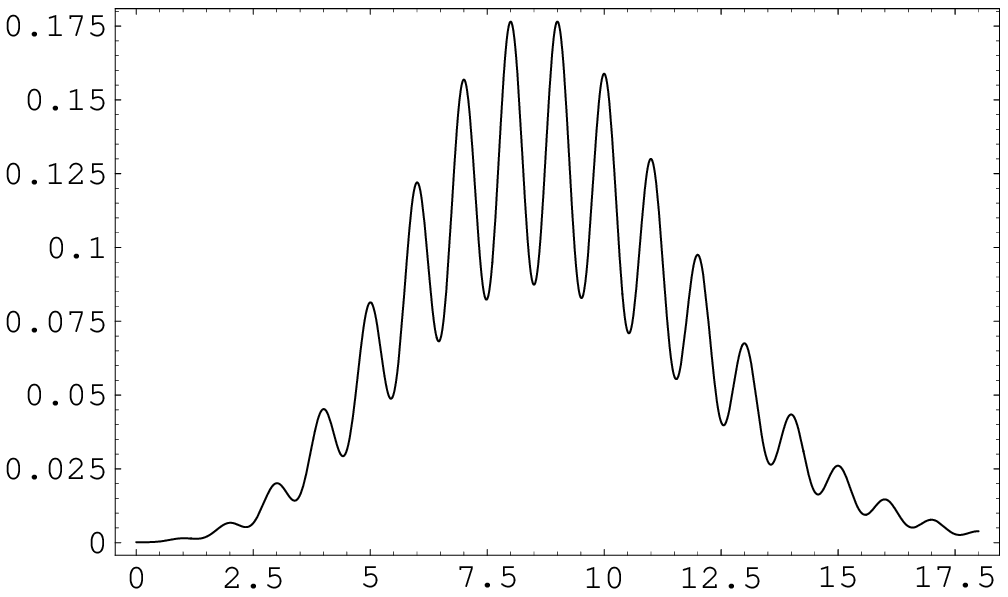}}}
\put(0,320){\makebox(40,20){\Large $P(n_m)$}}
\put(200,210){\makebox(20,20){\Large $n_m$}}
\put(90,170){\makebox(20,20){\Large (b)}}
\put(45,20){\makebox(300,180){
\includegraphics[width=10cm]{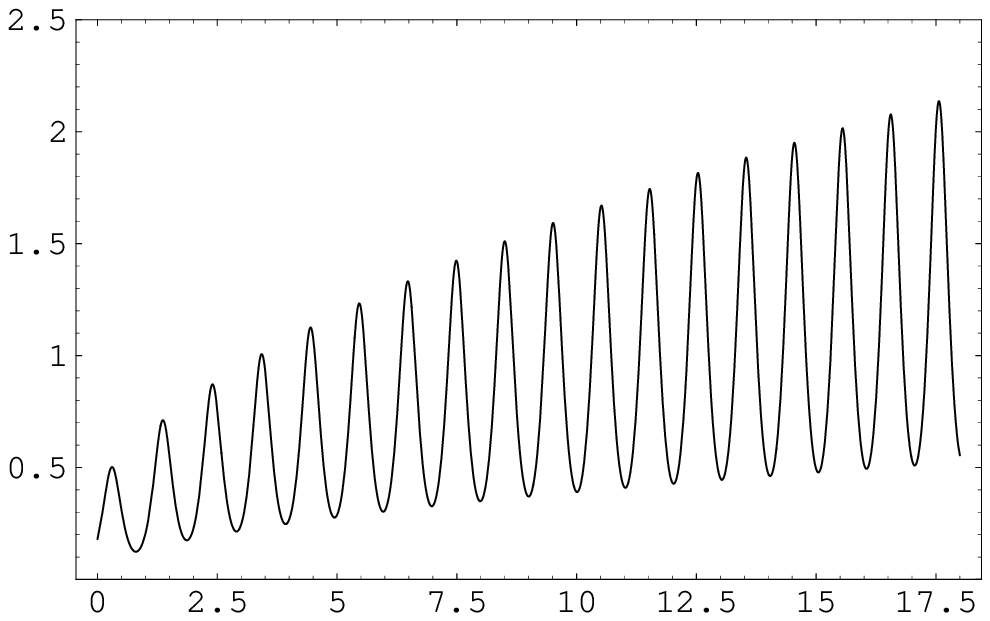}}}
\put(10,100){\makebox(40,20){\Large $\langle\hat{a}\rangle_f$}}
\put(200,0){\makebox(20,20){\Large $n_m$}}
\end{picture}
\caption{\label{stats} Photon number measurement statistics of a 
coherent state with 
an average amplitude of $\alpha=3$ at a photon number resolution of
$\delta\! n=0.3$. (a) shows the probability distribution over measurement
results and (b) shows the expectation value $\langle\hat{a}\rangle_f(n_m)$ 
after the measurement.}
\end{figure}
It reads
\begin{equation}
\langle \hat{a} \rangle_f(n_m) =
 \alpha \; \exp\left(-\frac{1}{8\delta\! n^2}\right)
\frac{\sum_n \frac{|\alpha|^{2n}}{n!} 
             \exp\left(-\frac{(n+\frac{1}{2}-n_m)^2}{2\delta\! n^2}\right)}
     {\sum_n \frac{|\alpha|^{2n}}{n!} 
             \exp\left(-\frac{(n-n_m)^2}{2\delta\! n^2}\right)}.
\end{equation}
This fraction of two sums of Gaussians has its maxima at half integer
photon numbers, where the denominator representing the measurement
probability is minimal. Figure \ref{stats} shows the probability 
distribution and the coherence after the measurement for $\alpha=3$
and $\delta\! n=0.3$. A direct comparison of the probability and
the coherence near $n_m=9$ is shown in figure \ref{detail}
It is obvious that integer measurement results $n_m$ are correlated
with strong decoherence and half integer results with weak decoherence.

\section{Nonclassical correlation}
\begin{figure}
\hspace{1.25cm}
\begin{picture}(345,200)
%\put(0,0){\framebox(345,200){}}
\put(40,20){\makebox(300,180){
\includegraphics[width=10cm]{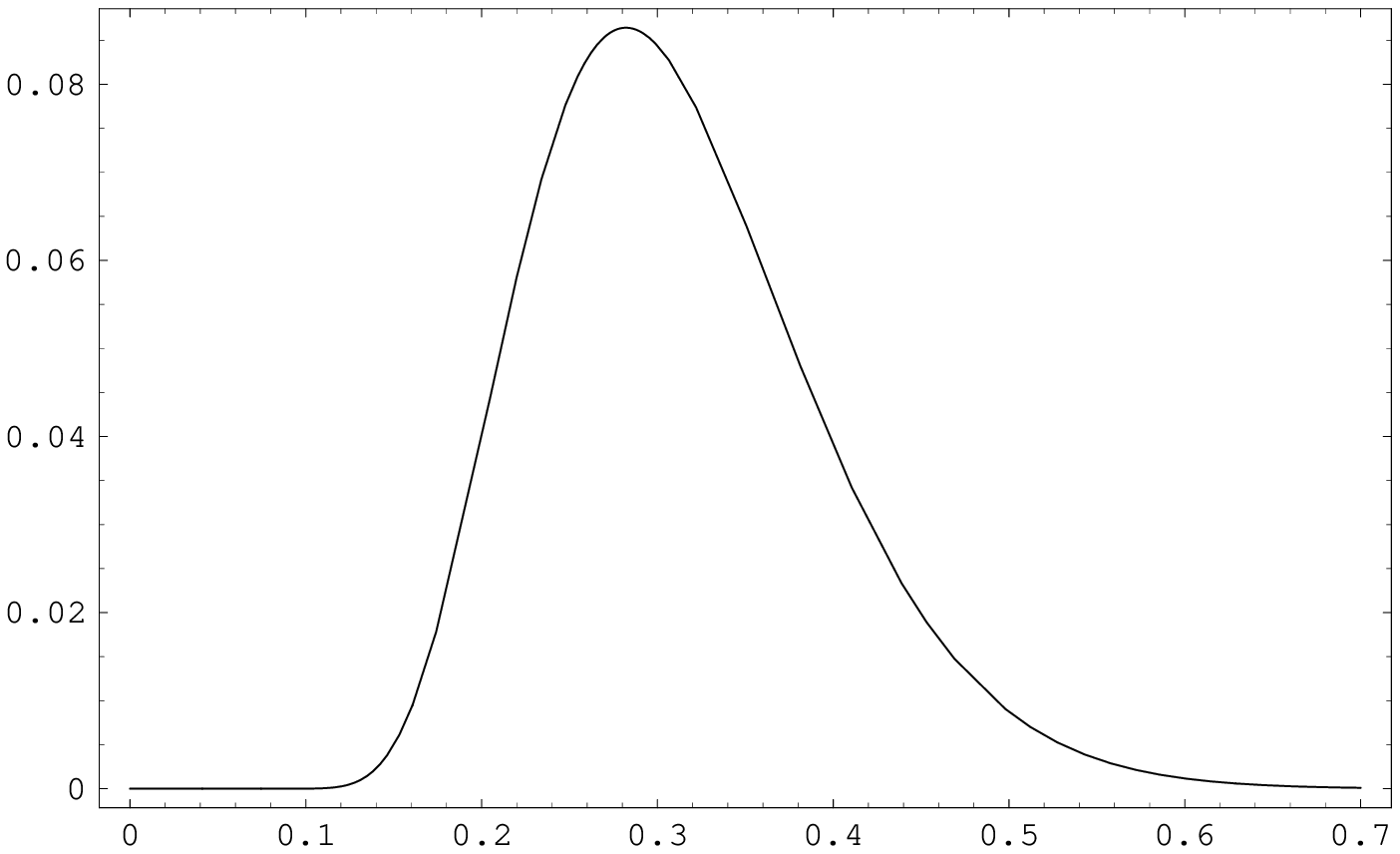}}}
\put(0,102){\makebox(40,50)
{\Large $-\frac{C(Q,\langle \hat{a}\rangle_f)}{\alpha}$}}
\put(190,0){\makebox(20,20){\Large $\delta\! n$}}
\end{picture}
\caption{\label{corrfig}
Normalized anticorrelation of the quantization 
$Q$ of the measurement result
$n_m$ and the coherence $\langle \hat{a} \rangle_f(n_m)$ after the measurement
as a function of measurement resolution $\delta \! n$.}
\end{figure}

If the quantization $Q$ of a measurement result $n_m$ is defined as
\begin{equation}
Q=\cos\left(2\pi n_m\right),
\end{equation}
then the correlation between quantization and coherence observed in the
finite resolution measurement of photon number may be expressed as
\begin{equation}
\label{eq:corr}
C(Q,\langle \hat{a}\rangle_f) =
  \int \left( Q(n_m) \; \langle \hat{a}\rangle_f(n_m)\right)d n_m 
- \left(\int Q(n_m) \; d n_m\right) \; 
  \left(\int \langle \hat{a}\rangle_f (n_m)\; d n_m\right).
\end{equation}
In the case of a coherent state, the averages of quantization
$Q$, of coherence $\langle \hat{a}\rangle_f$, and of their product
are given by
\begin{eqnarray}
\int Q(n_m) \; d n_m &=& \exp\left(-2\pi^2\delta\! n^2\right)
\\
\int \langle \hat{a}\rangle_f (n_m)\; d n_m &=& 
\exp\left(-\frac{1}{8 \delta\! n^2}\right) \alpha
\\
\int \left( Q(n_m) \; \langle \hat{a}\rangle_f(n_m)\right)d n_m 
&=& - \exp\left(-2\pi^2\delta\! n^2\right) 
      \exp\left(-\frac{1}{8 \delta\! n^2}\right) \alpha.
\end{eqnarray}
Quantization and coherence are therefore exactly anti-correlated,
with $C(Q,\langle \hat{a}\rangle_f)$ being equal to two times the
negative product of average quantization and average coherence,
\begin{eqnarray}
C(Q,\langle \hat{a}\rangle_f) &=& -2 \left(\int Q(n_m) \; d n_m\right) \; 
  \left(\int \langle \hat{a}\rangle_f (n_m)\; d n_m\right)
\nonumber \\ &=&
 - 2 \exp\left(-2\pi^2\delta\! n^2\right) 
      \exp\left(-\frac{1}{8 \delta\! n^2}\right) \alpha.
\end{eqnarray}
Figure \ref{corrfig} shows this anti-correlation as a function of measurement
resolution $\delta\! n$. Note that the anti-correlation is maximal near 
$\delta\! n=0.3$. For lower $\delta\! n$, decoherence reduces the average
value of $\langle \hat{a}\rangle_f (n_m)$ to zero. For higher $\delta\! n$,
quantization is not resolved in the measurement and the average value
of $Q$ is zero. 

The anti-correlation of quantization and coherence is a direct consequence
of quantum coherence. It originates from the fact that the operator $\hat{a}$
connects photon number states $\mid n \rangle$ with photon number states
$\mid n-1 \rangle$. This property means that $\hat{a}$ does not commute with
functions of $\hat{n}$. If the quantization operator $\hat{Q}$ is defined
as 
\begin{eqnarray}
\hat{Q} &=& \left(\cos(\pi\hat{n})\right)^2 - \left(\sin(\pi\hat{n})\right)^2
\nonumber \\
&=& (-1)^{2\hat{n}},
\end{eqnarray}
then the anti-correlation between quantization and coherence can be obtained
by ``splitting'' $\hat{Q}$ into two operators and inserting $\hat{a}$ in the
middle: 
\begin{eqnarray}
C(\hat{Q},\hat{a}) &=& \langle (-1)^{\hat{n}}\hat{a}(-1)^{\hat{n}} \rangle 
- \langle (-1)^{2\hat{n}} \rangle \langle \hat{a} \rangle
\nonumber \\ 
&=& - 2 \langle \hat{a} \rangle.
\end{eqnarray}
Of course, this operator ordering is only justified because the actual 
measurement first obtained $n_m$ and then $\langle \hat{a} \rangle$. 
The position of $\hat{a}$ between the parity operators $(-1)^{\hat{n}}$ 
is a consequence of this measurement context. 

\section{Context dependence of information}
In the example above, the measurement information obtained is not
information about the quantum state before the measurement, since this
state is already known to be a coherent state. Nevertheless, the quantum 
state does not provide sufficient information to describe the physical 
properties of the light field in every possible context. Therefore, relevant
new information about a physical property may be extracted
even from the pure state. There seems to be no classical analogy to this
process of extracting new information from a well defined state.
Possibly, the generalization of classical information
concepts to quantum mechanics is not as straightforward as the concepts of
entropy and qbits seem to suggest \cite{Ste00}. In particular, the measurement
interaction is responsible for selecting the relevant information. 
If information has been encoded in a different variable, then this
information becomes physically irrelevant and is lost. However, this loss
of information correponds to a smooth and continuous transition from
one type of context to the other, leaving some room for correlations such as
the one between quantization and coherence. 

In fact, it seems to be more
natural to consider situations in which the context is not given by a well
defined orthogonal set of states, but by a combination of non-commuting 
properties which could be described by a sequence of finite resolution
measurements. The problem of avoiding decoherence in quantum 
computation probably arises because it is difficult to limit the measurement 
context to precise measurements at well defined times. If the wrong type of
information leaks out, the artificial concepts of quantum information break
down. On the other hand, the full wealth of possibilities naturally inherent
in the quantum properties of physical properties may only be explored by
embracing the correlations between non-commuting variables. Specifically,
the nonclassical correlations expressed by context dependent operator 
ordering may provide a clearer understanding of the physical properties
behind such applications of quantum effects as quantum computation,
quantum teleportation, and quantum cryptography. 

\section{Conclusions}
%
%--------nonclassical correlations=new causalities
%
The uncertainty relations severely restrict the physical information
available about a quantum system. Nevertheless, new information may even 
be extracted from a pure state by measuring a previously unknown property.
This extraction of new information requires the loss of information
about the known system properties and therefore constitutes an
exchange of one type of physical information for another.
Yet this exchange process is more precise than the notion of uncertainty
would seem to suggest.

In the case discussed above, the coherent field information is gradually
exchanged for photon number information. However, an accidental 
observation of an ``incorrect'' half integer photon number does not result 
in a loss of information - instead the original phase information is retained. 
On the other hand, even the accidental observation of the ``correct'' integer
photon number does not provide more information on the system - instead,
there is an increased loss of phase coherence to compensate the precision of
the photon number measurement result. Thus the nonclassical correlation 
of quantization and coherence preserve the total information available about
the system while changing the physical context. 

It seems to be the major difference between classical physics and 
quantum physics 
that quantum physics establishes a fundamental relationship between
the information available about a system and the physical properties of
the system. In particular, bit like quantized information is only 
obtained if the property of coherence has been lost. The anti-correlation
of quantization and coherence demonstrates that quantization itself
is a property which depends on the measurement context and thus on the
information presently available about the light field. This ambiguity
in the physical properties appears to be the fundamental reason for 
the dualism between the physical and the mathematical principles
of quantum mechanics.

\section*{Acknowledgements}
One of us (HFH) would like to acknowledge support from the Japanese 
Society for the Promotion of Science, JSPS.

\vspace{1cm}
%=========================================================

%========================================================

\end{document}